# Low-level laser irradiation-induced changes in bovine spermatozoa

**Thiago Revers Dreyer**[a,c], **Adriano Felipe Perez Siqueira**[b], **Taciana Depra Magrini**[a], **Marcilio Nichi**[b], **Mayra E. O. D. Assumpção**[b], **Marcella Pecora Milazzotto**[a], **Herculano da Silva Martinho**[a]

[a]Universidade Federal do ABC, Centro de Ciências Naturais e Humanas, Rua Santa Adélia 166, Bangu, Santo André, SP, Brazil, 09210-170
[b]Universidade de São Paulo, Faculdade de Medicina Veterinária e Zootecnia, Av. Prof. Dr. Orlando Marques de Paiva, 87, Cidade Universitária, São Paulo, SP, Brazil, 05508-270
[c]Actual addres: Depto. Física e Biofísica, Instituto de Biociências de Botucatu, Unesp, Distrito de Rubião Junior s/n, Botucatu, SP, Brazil, 18618-970.

**Abstract**. Some biochemical and metabolic characteristics of bovine sperm cells irradiated with low-level laser were probed and correlated to the irradiation parameters. Bovine spermatozoa were exposed to spatially filtered light from a He-Ne laser (633 nm) with fluences from 150 to 600 mJ/cm². General biochemical alterations were assessed by vibrational spectroscopy (micro-Fourier-Transform Infrared technique). The mitochondrial membrane potential was assessed by flow cytometry. Lipid peroxidation due to reactive oxygen species (ROS) production was assessed by thiobarbituric acid reactive substances (TBARS) assay. Sperm morphologic characteristics changes were probed by atomic force microscopy (AFM) technique. LLLI was able to induce, at different fluences, changes in lipid, DNA and protein contents. Moreover, LLLI also induced changes in TBARS production, suggesting an irradiation effect on lipid peroxidation. Acrosomal reaction induction evidences was found based on changes in protein structure (Amides I and II bands) and AFM characterization. The most interesting result, however, was the changes in methyl and DNA groups, which strongly suggests that LLLI can modify DNA methylation patterns, important for pregnancy establishment. The observed changes in the above-cited parameters have direct impact on sperm quality. The low-level laser irradiation has the potential to improve the fertilizing capacity of bovine sperm cells.

# 1   Introduction

The sperm cell is an interesting model to study the light-tissue interactions since it has few structures and is a very specialized cell. The sperm conservation and manipulation has great economic impact in many fields, such as the preservation of endangered species and developments in farming, as well as in human reproduction. In this sense, any system that modulates sperm function can be applied to improve the strategies used in sperm conservation and manipulation[1].

The development of assisted reproductive technologies, like *in vitro* fertilization and intracytoplasmic sperm injection has provided relevant contributions to animal breeding programs[2]. These biotechniques, when applied to cattle, in most of the cases are conducted with frozen semen. The cryopreservation process, although very studied, still leads to damage of the sperm cell, like DNA fragmentation[3], loss of motility[4], plasma and acrosomal membrane injury and changes in chromatin structure[5], reducing sperm fertility.

The usual way[6–9] to prevent its damage and avoid poor quality of frozen semen is using cryopreservers in the freezing process. However, the paradox is that cryopreservers themselves can have a toxic effect on sperm as membrane destabilization, protein and enzyme denaturation. This effect is related directly to the concentration used and the time of cell exposure[10,11]. Recently the low-level laser irradiation (LLLI) has been applied to frozen sperm to improve its quality. The results reported on literature are very promising[12–17].

Corral-Baqués[12] showed that irradiation of canine sperm with diode laser at 655 nm at fluences of 4, 6 and 10 J/cm² improves velocity and linear coefficient of sperm. It was also reported an increase in the sperm fertilizing capacity[13] and motility[14], in addition to the increase of calcium ions amount in the irradiated sperm cells[15].

It has been reported that the irradiation effects depend on fluences, laser wavelength, and animal species. Stored turkey spermatozoa irradiated with 3.96 J/cm$^2$ presented an increased longevity[16], while for bovine it was observed a decrease in mortality with increasing fluences up to 16 J/cm$^2$ [17]. Moreover, fish sperm had an increase in motility and fertility post-irradiation with red (660 nm) and white light (400-800 nm), while in ram sperm motility and fertility were increased just with red light, under the same conditions of irradiation. However, irradiation with violet and ultraviolet sources (360 and 294 nm, respectively) induced a decrease in motility and fertility from both animal species[14].

It is important to stress that in spite of many investigations on the subject, the fundamental molecular and cellular mechanisms responsible for signal transduction from the photons that are incident on the cells to the biological effects that take place in the illuminated tissues is still uncertain[18].

Most of the methods routinely used to evaluate seminal parameters and considered essential to determine sperm cell fertility are based on general physical characteristics of the sperm cells as morphology, motility patterns and integrity of organelles[19]. However, the results of these evaluations are discrepant because they depend on the technique and the conditions that they are performed[4,20]. Moreover there is still a difficulty to standardize the techniques, since different research groups use incomparable analytical methodologies. Moreover some analysts have a poor distinction among sperm velocity patterns, making internal and external quality control very difficult to be done[21,22].

Recently, non-invasive or minimally invasive biophotonic techniques have been applied to explore biochemical variations in biological tissues. These techniques are able to provide more accurate data related to the metabolic status of the cell. The Fourier-transform infrared spectroscopy (FTIR) is one of them. FTIR spectroscopy is a technique which provides pieces of information on molecular vibrational of a sample giving a detailed fingerprint of different bonds, functional groups, and conformations of molecules[23]. It is a non-invasive and non-destructive method to probe the molecular/biochemical composition of a given sample. Thus it could be a usual quantitative tool enabling one to estimate the metabolic fingerprint of cells like spermatozoa[24]. The main goal of the present work was to assess the biochemical and metabolic alterations of bovine sperm cells irradiated with LLLI aiming to contribute to the understanding of the light-tissue interaction in this system.

## 2 Materials and methods

### 2.1 Sperm preparation

Bovine sperm straws were thawed in water bath at 37°C for 30 seconds and centrifuged (9,000 xg, 5 min). The supernatant was discharged and the sperm concentration was measured in a Neubauer chamber. The final concentration was adjusted to $7 \times 10^6$ sperm cells/mL in TALP medium (Tyrode's salt solution with albumen, lactate and pyruvate) and the final volume was divided in two Petri dishes, one to be irradiated and the other to be kept in the same conditions but without red light irradiation (control). Analyses were performed just after irradiation (PI) and after 30 minutes, time called as incubation period (I). During this period, sperm samples were kept at 38.5°C, 5% of $CO_2$ and high humidity.

After irradiation, samples were washed by centrifugation (9,000 x g for 5 min) and the pellet was resuspended in saline solution (NaCl, 0.9%) two times. The last centrifugation was performed for 20 minutes. A volume (5 μL) of the final supernatant and the resulting pellet resuspended in 5 μL of its supernatant were transferred to a platinum recovered surface plate for FTIR analysis. The remaining sample was used for mitochondrial membrane potential and lipid peroxidation (tiobarbituric acid reaction species, TBARS) biochemical analysis.

*2.2 Irradiation parameters*

Irradiation was performed with a He-Ne laser (TEM00, Unilaser, Brazil) at a wavelength of 633 nm and an output power of 24 mW being adjusted to the powers of 5, 7.5 and 10 mW through the use of filters. The irradiation setup details were reported by Magrini and coworkers[23]. Semen samples were irradiated with laser fluences of 150, 230, 300, 450 and 600 mJ/cm² according to the power and exposure times listed in Table 1. The fluences of 300 mJ/cm² were obtained by two different conditions of irradiation (power and irradiation time), and are identified and differentiated by the superscripts "ε" and "φ".

**Table 1** Irradiation parameters of the experiment.

| Fluence (mJ/cm²) | 150 | 230 | 300$^ε$ | 300$^φ$ | 450 | 600 |
|---|---|---|---|---|---|---|
| Power (mW) | 5 | 7.5 | 10 | 5 | 7.5 | 10 |
| Irradiation time (min) | 5 | 5 | 5 | 10 | 10 | 10 |

## 2.3 Micro FTIR spectroscopy

A Varian 610 FTIR Micro-spectrometer was used in reflectance mode, with a spectral resolution of 4 cm$^{-1}$, 400 scans per background and 200 scans in 4 different points of each sample. Each spectrum was manually baseline corrected with Fityk software[25] and the average spectrum was computed. If the average spectrum of the supernatant presented some band with intensity higher than 5% of the sperm spectrum, the region of the sample spectrum was excluded from the analysis.

## 2.4 Mitochondrial membrane potential

Mitochondrial membrane potential was determined with the lipophilic cationic dye 5,5',6,6'-tetrachloro-1,1',3,3'-tetraethylbenzimidazolcarbocyanine iodide (JC-1) according to the instructions of the manufacturer (Molecular Probes, USA). Briefly, 1.05x10$^6$ cells were incubated for 10 min at 25°C with JC-1 solution. The sperm populations with high mitochondrial membrane potential (HMMP), intermediate (IMMP) and low mitochondrial membrane potential (LMMP) were determined by flow cytometer in a Guava EasyCyte™ Mini System (Millipore, USA).

## 2.5 TBARS test

The method is based on the reaction of two molecules of tiobarbituric acid (TBA) with a molecule of malondialdehyde (MDA), which is the main product of lipid peroxidation, in high temperature and low pH, resulting in a pink chromogen that is quantified in a spectrophotometer. Two methodologies were performed: natural and induced lipid peroxidation. The first one aims to verify the oxidative stress to which the sample had already been exposed to, while the second indicates the susceptibility of the sperm cells to the oxidative stress.

The natural and induced lipid peroxidation quantification was assessed according to Nichi[26]. The induction of lipid peroxidation was performed by the addition of iron sulfate (100 μL, 4mM) and ascorbic acid (100 μL, 20 mM) to 186 μL of PBS and 214 μL of semen sample followed by incubation at 37°C for 2 hours.

*2.6 Atomic force microscopy (AFM) imaging*

The surface of an extra-smooth silicon slide was cleaned with ethanol and air dried. Then 50 μL of spermatozoa suspension were diluted in 1:10 phosphate buffered saline (PBS), placed on a silicon slide and air-dried. The silicon slide was immediately placed into sample holder and finally scanned by AFM[27]. Scanning of spermatozoa cells was performed in contact mode by using a commercial AFM (5500 Atomic Force Microscope, Agilent Technologies) in air at room temperature by using Nanosensor probes (PPP-Cont, Pointprobe Plus – tip, Nanosensors) with 0.2 N/m elastic constant.

*2.7 Statistical analysis*

The baseline-corrected FTIR spectra were vector normalized and analyzed by Principal Component Analysis (PCA). The FTIR data were grouped according their characteristic vibrational bands. These bands were separately used for the Principal Components (PC) calculations by computing the covariance matrix. They were used as predictors in binary logistic regression (BLR) model to differentiate the irradiated groups from the corresponding controls.

The number of PCs used in the BLR was determined by the analysis of the eigenvalue vs component number plot (scree plot) and by their contribution to description of the data (higher than 0.5%).

BLR was used to classify the spectra in groups according to its quantitative measurements and returned the percentage of concordant pairs. The percentage threshold of concordant pairs was defined by the control groups of the samples irradiated by 5 and 10 minutes, since they presented higher similarity. The comparisons among the irradiated groups with different fluences and their controls that presented difference by the PCA analysis were done by the integrated areas under their average normalized spectra.

Comparison between control and irradiated data for TBARS and flow cytometry were evaluated by $t$-Student test and Mann-Whitney for parametric and nonparametric variables, respectively (SAS System for Windows (SAS Institute Inc, Cary, NC, USA)). For different fluences, the analysis was performed by Least Significant Differences (LSD) test. Correlations among the variables were calculated by Pearson and Spearman tests respectively for parametric and nonparametric variables, being described by its correlation coefficient ($r$) and significance level ($p$).

## 3   Results and discussion

Figure 1 presents the average FTIR spectra of a semen pellet compared to supernatant and saline solution. The corresponding vibrational band assignment was based on ref. 28 is presented on Table 2.

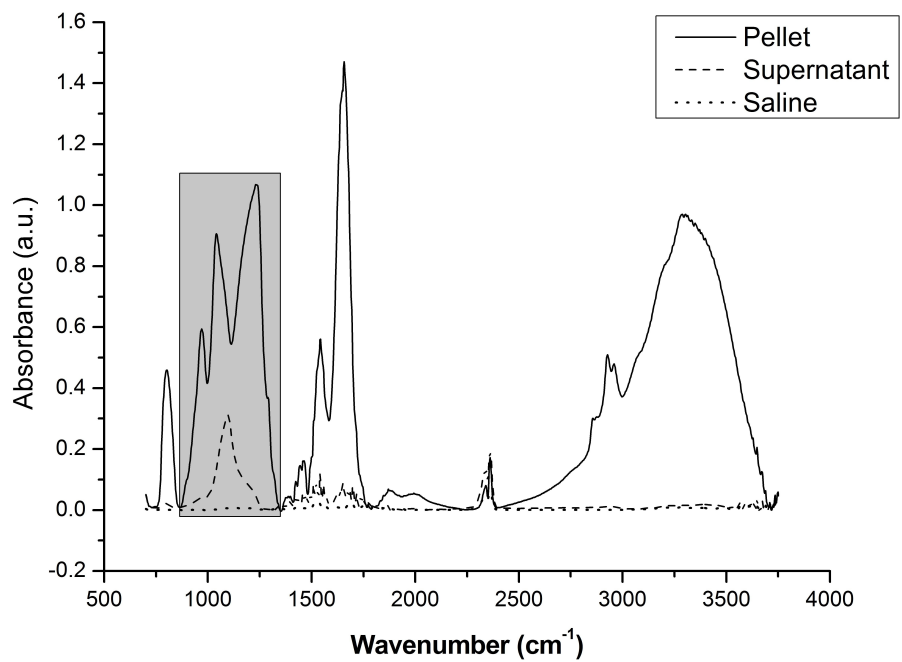

**Fig. 1** Representative analysis of each sample spectra validation. The solid line is the mean spectrum of the sample constituted mainly by spermatozoa (Pellet), dashed line is the average spectrum of the supernatant (Supernatant) and the dot line is the average spectrum of saline buffer (Saline). In this case the shaded region was removed from the statistical analysis, since the supernatant spectrum intensity of the bands in this region was higher than the threshold limit (5%).

**Table 2** FTIR general vibrational band assignment[28].

| Spectral range of frequencies (cm$^{-1}$) | Biological assignment |
|---|---|
| 760 – 860 | DNA |
| 1,584 – 1,766 | Amide I |
| 1,486-1,584 | Amide II |
| 1,410-1,488 | Methyl groups |
| 2,821 – 3,600 | Lipids and water |

The DNA, proteins and lipids presented high variation after PC analysis (data not shown). Based on the analysis of the scree plot and the representativeness of the data of the PCs, the first 3 PCs were used to calculate the percentage of concordant pairs in the BLR. The threshold of statistical difference between the irradiated groups and its corresponding controls was 66.4% of concordant pairs to DNA region (760-860 cm$^{-1}$), 65.5% to proteins (1,410-1,766 cm$^{-1}$) and 61% to lipids (2,821-3,600 cm$^{-1}$). The same procedure related to BLR was done with the biochemical components of all groups irradiated and their controls for each fluency. Almost all biochemical groups presented the number of concordant pairs above the threshold. Due to this its were considered as having statistically differences between irradiated and controls groups. The only groups that showed number of concordant pairs below the threshold were DNA of incubated (I) samples irradiated with 150 mJ/cm² and 600 mJ/cm² with their respective controls and; DNA and lipids for the groups analyzed just after irradiation (PI) with 230 mJ/cm² and 600 mJ/cm², and therefore presented no statistical differences between irradiated and control groups (Table 3).

**Table 3** Percentage of concordant pairs obtained in BLR to samples irradiated with different fluences and their respective controls. Cells filled in gray present higher percentage of concordant pairs than threshold and were taken to analyses by the integrated areas under the spectral curves.

| Fluence (mJ/cm²) | PCs | PI | | | I | | |
|---|---|---|---|---|---|---|---|
| | | DNA | Protein | Lipid | DNA | Protein | Lipid |
| 150 | 3PCs | 78.2 | 73.6 | 79.5 | 64.1 | 70.5 | 68 |
| | PC1+PC2 | 75 | 71.8 | 81.9 | 63.9 | 66.7 | 68.2 |
| | PC1+PC3 | 67.2 | 74 | 79.6 | 61 | 68.5 | 63.2 |
| | PC2+PC3 | 71.9 | 56.6 | 68 | 64.1 | 69.8 | 68.2 |
| 230 | 3PCs | 65.4 | 76.3 | 69.8 | 65.1 | 84.2 | 73 |
| | PC1+PC2 | 57.2 | 62.7 | 68.1 | 50.5 | 54.5 | 73.4 |
| | PC1+PC3 | 64.9 | 65.4 | 72.3 | 66.8 | 84.2 | 59.4 |
| | PC2+PC3 | 55 | 69.1 | 60.4 | 61.7 | 82.4 | 71.7 |
| 300 | 3PCs | 78.1 | 80.6 | 85.4 | 75.6 | 79.5 | 69.6 |
| | PC1+PC2 | 76.8 | 79.6 | 85.2 | 68.4 | 67.9 | 68.7 |
| | PC1+PC3 | 73.8 | 80.9 | 83.9 | 70.9 | 74.6 | 51.3 |
| | PC2+PC3 | 66.9 | 51.6 | 65.9 | 72.3 | 79.1 | 70 |
| 300* | 3PCs | 69.5 | 72 | 95.9 | 70.8 | 77.6 | 92.9 |
| | PC1+PC2 | 69 | 63.9 | 96 | 71.4 | 69 | 92.5 |
| | PC1+PC3 | 55.3 | 68.8 | 52.8 | 68.2 | 76.2 | 69.5 |
| | PC2+PC3 | 69.5 | 71 | 95.9 | 70.4 | 78.1 | 92.9 |
| 450 | 3PCs | 76.3 | 87.8 | 82.8 | 70.2 | 69.1 | 60.9 |
| | PC1+PC2 | 73.2 | 86.1 | 74.4 | 62.7 | 68.5 | 62.3 |
| | PC1+PC3 | 76.1 | 82.7 | 83 | 69.1 | 69.3 | 58.9 |
| | PC2+PC3 | 52.6 | 64.5 | 69.6 | 59.6 | 58,1 | 56.3 |
| | 3PCs | 73 | 83.8 | 57.6 | 66.2 | 66.4 | 67.4 |
| 600 | PC1+PC2 | 70.8 | 81.7 | 58 | 61.6 | 66.9 | 60.1 |
| | PC1+PC3 | 68.9 | 58.6 | 53.1 | 64.7 | 60,4 | 67.3 |
| | PC2+PC3 | 73.9 | 83.6 | 57.4 | 64.3 | 68.6 | 69.5 |

After PCA analyses, protein region (1,410-1,766 cm$^{-1}$, see Fig. 1) were deconvoluted by its main components (Amide I (1,584-1,766 cm$^{-1}$), Amide II (1,486-1,584 cm$^{-1}$) and methyl groups of biomolecules (1,410-1,486 cm$^{-1}$)). The biochemical groups that presented differences in the BLR analysis (number of concordant pairs above the threshold) had their original baseline-subtracted spectra normalized by its integrated area under its spectral curves and the statistical analysis of the comparisons are described on Table 4, in which the empty cells showed no statistical difference between the groups.

**Table 4** Statistical analysis (p-value, t-Student) of the integrated areas under the spectral curves for the different components. Lines with gray background in the table correspond to the PI condition.

| Fluence (mJ/cm²) | Integrated areas under the curves in the spectral region | | | | | |
|---|---|---|---|---|---|---|
| | DNA | | | CH$_3$ of proteins | | |
| | Control | Irradiated | *p* | Control | Irradiated | *p* |
| 150 | 0.0336±0.0143 | 0.0405±0.0103 | 0.01 | 0.0167±0.0055 | 0.0148±0.0027 | 0.035 |
| 150 | 0.0400±0.0107 | 0.0454±0.0116 | 0.035 | 0.0145±0.0020 | 0.0182±0.0065 | 0.002 |
| 300ε | | | | 0.0167±0.0055 | 0.0145±0.0025 | 0.013 |
| 300ε | 0.0400±0.0107 | 0.0327±0.0088 | 0.002 | 0.0145±0.0020 | 0.0159±0.0026 | 0.017 |
| 300φ | | | | 0.0168±0.0040 | 0.0130±0.0021 | 0.001 |
| 450 | | | | 0.0168±0.0040 | 0.0140±0.0022 | 0.001 |
| 450 | | | | 0.0146±0.0028 | 0.0161±0.0024 | 0.04 |
| 600 | 0.0233±0.0094 | 0.0304±0.0109 | 0.005 | 0.0168±0.0040 | 0.0135±0.0017 | 0.001 |
| 600 | 0.0296±0.0157 | 0.0377±0.0147 | 0.016 | | | |
| | Amide II | | | Amide I | | |
| | Control | Irradiated | *p* | Control | Irradiated | *p* |
| 150 | | | | 0.1434±0.0124 | 0.1374±0.0125 | 0.032 |
| 230 | | | | 0.1464±0.0186 | 0.1396±0.0088 | 0.028 |
| 300ε | | | | 0.1464±0.0186 | 0.1560±0.0110 | 0.005 |
| 300ε | 0.0419±0.0109 | 0.0521±0.0103 | 0.001 | 0.1434±0.0124 | 0.1567±0.0115 | 0.001 |
| 300φ | 0.0620±0.0124 | 0.0477±0.0133 | 0.001 | 0.1590±0.0142 | 0.1432±0.0135 | 0.001 |
| 300^φ |  |  |  | 0.1480±0.0142 | 0.1364±0.0136 | 0.002 |
| 450 |  |  |  | 0.1480±0.0142 | 0.1573±0.0137 | 0.022 |
| 600 | 0.0620±0.0124 | 0.0472±0.0093 | 0.001 | 0.1590±0.0142 | 0.1497±0.0095 | 0.001 |
| 600 | 0.0495±0.0152 | 0.0421±0.0136 | 0.02 |  |  |  |
|  | **Total proteins** | | | **Lipids** | | |
|  | **Control** | **Irradiated** | *p* | **Control** | **Irradiated** | *p* |
| 230 |  |  |  | 0.1041±0.0066 | 0.0963±0.0049 | 0.001 |
| 300^ε | 0.1998±0.0229 | 0.2247±0.0223 | 0.001 | 0.1041±0.0066 | 0.1001±0.0080 | 0.028 |
| 300^φ | 0.2378±0.0282 | 0.2039±0.0257 | 0.001 | 0.1004±0.0145 | 0.0908±0.0120 | 0.001 |
| 300^φ | 0.2121±0.0294 | 0.1969±0.0244 | 0.022 | 0.1026±0.0091 | 0.0872±0.0102 | 0.001 |
| 450 |  |  |  | 0.1004±0.0145 | 0.0937±0.0050 | 0.01 |
| 450 |  |  |  | 0.1026±0.0091 | 0.0974±0.0062 | 0.01 |
| 600 | 0.2378±0.0282 | 0.2104±0.0176 | 0.001 |  |  |  |
| 600 | 0.2121±0.0294 | 0.2001±0.0223 | 0.033 |  |  |  |

Samples right after the irradiation (PI) with fluence of 150 and 600 mJ/cm² presented higher DNA integrated areas than their controls (Table 4). After incubation period, DNA from irradiated samples did not present differences from the control ones suggesting some reversible biochemical change induced by irradiation as, for an example, conformational changes between protamines and DNA. A change like this could be reflected in DNA integrity, which is related to sperm quality[8] and to embryonic development[4].

Interestingly, the amount of methyl groups also varied from control and irradiated samples. Except for 230 mJ/cm² all the groups had a decrease in methyl groups area when compared to their controls right after the irradiation, followed by an increase in this area after the incubation (Table 4). These results are important since the sperm cell's methylation pattern matched that of the active embryo cell at the zygote genome activation, while the egg cell needed to be reprogrammed before it reached this state. This means that at fertilization, the DNA from the sperm is more prone for development than that of the egg cell[29]. The irradiation seems to be able to induce methylation changes, which could be important to be applied at nuclear reprogramming studies. Moreover, after irradiation and incubation the laser induced an increase in methyl groups. This kind of modulation can be very usefull since the DNA methylation levels are correlated with pregnancy rates. In human, sperm DNA methylation was not related with fertilization and embryo development rates, however, higher methylation levels were related to higher rates of pregnancy establishment[30].

We showed by FTIR that Amide I and Amide II regions, which are related with conformational structure of proteins, presented high variation due to irradiation for all groups. The ability of LLLI to induce these conformational changes is very interesting since several physiological processes related to sperm maturation and acquisition of fertilization ability are linked to protein changes. As an example of this change is the acrosomal reaction, which promotes the recognition, adhesion and fusion of the sperm cell with the oocyte. There are several reports about the sperm morphologic characteristics by the use of AFM[27,31,32] technique which is able to provide information in tridimentional array of the sperm cell in its native environment. It was already shown[32] that the acrosome reacted sperm cell presented narrower thickness in the acrosome region. We observed similar results .Figure 2 shows tridimensional topography images of sperm cells just after trawling (Fig. 2a ), PI with fluence of 450 mJ/cm$^2$ (Fig. 2b), and I with same fluency (Fig. 2c). It is clear the alterations in the height and volume of the equatorial region of the sperm cell with irradiation, as indicated by the arrow in the Fig. 2c). As reviewed by Fénichel[33], this process happens due to externalization of ligand proteins, protein migration through the plasma membrane and conformational changes of pre-existing membrane proteins. These conformational changes were already described for HSP60, for example, facilitating the formation of a functional zona pellucida receptor complex on the surface of spermatozoa[34]. The ability to change the protein conformation raises the LLLI to a powerful tool to increase the sperm fertility.

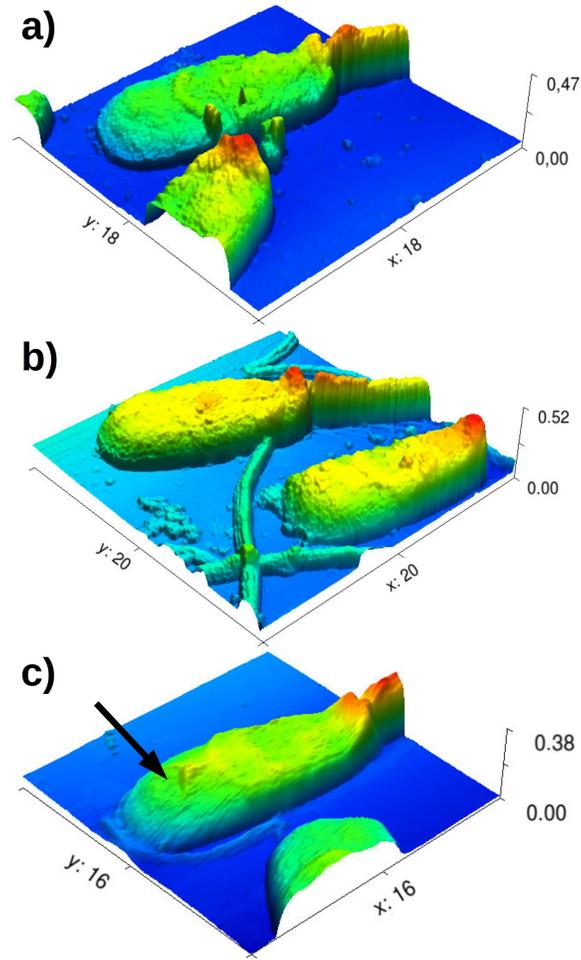

**Fig. 2** Tridimenstional topography of sperm cells just after thawing (a), PI (b) and I (c) with fluence of 450 mJ/cm² obtanied by AFM. The arrow in c) indicates the reduction volume of the acrosomal region.

Despite the importance of the biomolecules described until now, the amount and distribution of lipids in the sperm plasma membrane play a key role in modulating signaling pathways in this cell[35]. Changes in cholesterol and other lipid molecules alter during capacitation, determining the fecundation ability of the cell. In vitro, capacitation can be induced by albumin, which decreases the cholesterol/phospholipid ratio[36]. In our study, the ability to modulate these molecules with laser were observed for 230, 300 and 450 mJ/cm$^2$, mainly after the incubation period, suggesting a long term effect of the radiation on these molecules. Besides that, no changes were observed in relation to lipid peroxidation (TBARS test) for any fluence when compared to its control (Table 5 and 6).

**Table 5** Fluence effects obtained by the irradiation with times of 5 minutes with laser powers of 5, 7.5 and 10 mW for the variables analyzed by TBARS assays (TBARS I and TBARS E) and by flow cytometry (High MMP, Low MMP and Intermediate MMP).

|  | 0 mJ/cm$^2$ | | 150 mJ/cm$^2$ | | 230 mJ/cm$^2$ | | 300 mJ/cm$^2$ | |
| --- | --- | --- | --- | --- | --- | --- | --- | --- |
| TBARS I (ng TBARS/mL semen) | 741.85 | 45.55 | 694.36 | 31.51 | 780.14 | 44.13 | 801.19 | 63.71 |
| TBARS E (ng TBARS/mL semen) | 435.49 | 8.29$^a$ | 440.77 | 7.45$^a$ | 450.04 | 1.42 | 420.15 | 12.95 |
| High MMP (%) | 33.11 | 1.91 | 30.01 | 2.79 | 31.10 | 3.26 | 31.50 | 2.86 |
| Low MMP (%) | 41.36 | 2.08 | 35.85 | 3.99 | 38.24 | 3.79 | 41.83 | 3.22 |
| Intermediate MMP (%) | 26.40 | 1.44 | 34.15 | 2.79 | 30.64 | 2.85$^a$ | 26.66 | 2.93 |

**Table 6** Fluence effects obtained by the irradiation with 10 minutes with laser power of 5, 7.5 and 10 mW to the variables analyzed by TBARS assay (TBARS I and TBARS E) and by flow cytometry (high MMP, low MMP and intermediate MMP).

|  | 0 mJ/cm$^2$ | | 300 mJ/cm$^2$ | | 450 mJ/cm$^2$ | | 600 mJ/cm$^2$ | |
| --- | --- | --- | --- | --- | --- | --- | --- | --- |
| TBARS I (ng TBARS/mL semen) | 813.27 | 48.86$^a$ | 898.44 | 82.20 | 813.06 | 42.61 | 671.42 | 39.92 |
| TBARS E (ng TBARS/mL semen) | 444.29 | 5.26 | 448.82 | 1.35 | 491.50 | 51.62 | 466.43 | 49.08 |
| High MMP (%) | 31.62 | 1.59 | 28.70 | 2.19 | 30.50 | 3.3 | 29.25 | 3.17 |
| Low MMP (%) | 43.36 | 1.94 | 44.08 | 2.69 | 38.56 | 4.82 | 40.18 | 3.38 |
| Intermediate MMP (%) | 25.15 | 1.05 | 27.22 | 1.47$^a$ | 30.94 | 2.23 | 30.60 | 2.09 |

These data corroborate with the lack of changes in mitochondrial membrane potential induced by the laser, which could induce a higher generation of ROS and, consequently, lipid peroxidation. Kujawa[34] described that near-infrared low-intensity laser radiation (10-12 J/cm$^2$) changes ATPase activities in red blood cells but does nI.ot alter integral parameters as cell stability and membrane lipid peroxidation level. In contrast, for patelets, Trofimov[35] used low intensity He-Ne laser (6 J/cm$^2$) promoting metabolic rearrangements in lipids and activation of lipid-dependent cell systems. These data suggest that the effect of radiation on plasma membrane lipids seems to vary among different cell types, lasers aI.nd doses and can be modulate by laser irradiation. Moreover, the parameters of laser irradiation used in this study were efficient in alter the sperm cell lipid content without alter metabolic pathways that could interfere on this content in a negative way.

## 4  Conclusions

Based on our results we can conclude that LLLI were capable of induce metabolic and structural changes on bovine sperm cell. It was shown a possible conformational changes between DNA and protamines, induction in the methylation patterns which could be important in sperm quality. That was also shown alteration in the morphology of the sperm cell after the incubation period, which could be related to acrosomal reaction, which also could be related to the modulation of the lipids content of the sperm cell. However, further work is need since the cascade of events related to the interaction light-sperm cells are yet not completely known. By the discovery of these mechanisms the improvement of fertilizing capacity through electromagnetic radiation will be achieved.

## 5  Acknowledgments

We would like to thank the sponsorship of CAPES, FAPESP (2009/51630-9, 2011/06618-5) and CNPq. We are also grateful for the Multiuser Central Facilities of UFABC for the experimental support.

# 6. References

1. M. I. Corral-Baqués et al., "The effect of low-level laser irradiation on dog spermatozoa motility is dependent on laser output power.," *Lasers Med. Sci.* **24**(5), 703–713 (2009).

2. T. C. D. Mundim et al., "Changes in gene expression profiles of bovine embryos produced in vitro, by natural ovulation, or hormonal superstimulation.," *Genet. Mol. Res.* **8**(4), 1398–1407 (2009).

3. L. Fraser and J. Strzezek, "Effect of different procedures of ejaculate collection, extenders and packages on DNA integrity of boar spermatozoa following freezing-thawing.," *Anim. Reprod. Sci.* **99**(3-4), 317–329 (2007).

4. P. F. Watson, "Recent developments and concepts in the cryopreservation of spermatozoa and the assessment of their post-thawing function.," *Reprod. Fertil. Dev.* **7**(4), 871–891 (1995).

5. A. Januskauskas, A. Johannisson, and H. Rodriguez-Martinez, "Assessment of sperm quality through fluorometry and sperm chromatin structure assay in relation to field fertility of frozen-thawed semen from Swedish AI bulls," *Theriogenology* **55**(4), 947–961 (2001).

6. S. Sariözkan et al., "The influence of cysteine and taurine on microscopic-oxidative stress parameters and fertilizing ability of bull semen following cryopreservation.," *Cryobiology* **58**(2), 134–138 (2009).

7.	M. N. Bucak et al., "Effects of antioxidants on post-thawed bovine sperm and oxidative stress parameters: antioxidants protect DNA integrity against cryodamage.," *Cryobiology* **61**(3), 248–253 (2010).

8.	P. B. Tuncer et al., "The effect of cysteine and glutathione on sperm and oxidative stress parameters of post-thawed bull semen.," *Cryobiology* **61**(3), 303–307 (2010) .

9.	M. J. McCarthy and S. a Meyers, "Antioxidant treatment in the absence of exogenous lipids and proteins protects rhesus macaque sperm from cryopreservation-induced cell membrane damage.," *Theriogenology* **76**(1), 168–176, Elsevier Inc. (2011).

10.	M. P. Rosato and N. Iaffaldano, "Cryopreservation of rabbit semen: comparing the effects of different cryoprotectants, cryoprotectant-free vitrification, and the use of albumin plus osmoprotectants on sperm survival and fertility after standard vapor freezing and vitrification.," *Theriogenology* **79**(3), 508–516 (2013).

11.	J. E. Swain and G. D. Smith, "Fertility Cryopreservation," in *Fertil. Cryopreserv.*, 1st ed., R.-C. Chian and P. Quinn, Eds., pp. 24–38, Cambridge University Press, New York (2010).

12.	M. I. Corral-Baqués et al., "Effect of 655-nm diode laser on dog sperm motility.," *Lasers Med. Sci.* **20**(1), 28–34 (2005).

13.	N. Cohen et al., "Light irradiation of mouse spermatozoa: stimulation of in vitro fertilization and calcium signals.," *Photochem. Photobiol.* **68**(3), 407–413 (1998).

14.	T. Zan-Bar et al., "Influence of visible light and ultraviolet irradiation on motility and fertility of mammalian and fish sperm.," *Photomed. Laser Surg.* **23**(6), 549–555 (2005) .

15. R. Lubart et al., "Effect of light on calcium transport in bull sperm cells.," *J. Photochem. Photobiol. B.* **15**(4), 337–341 (1992).

16. N. Iaffaldano et al., "Improvement of stored turkey semen quality as a result of He-Ne laser irradiation.," *Anim. Reprod. Sci.* **85**(3-4), 317–325 (2005).

17. J. M. Ocaña-Quero et al., "Biological effects of helium-neon (He-Ne) laser irradiation on acrosome reaction in bull sperm cells.," *J. Photochem. Photobiol. B.* **40**(3), 294–298 (1997).

18. A. C.-H. Chen et al., "Low-level laser therapy activates NF-kB via generation of reactive oxygen species in mouse embryonic fibroblasts.," *PLoS One* **6**(7), e22453 (2011).

19. P. F. N. Silva and B. M. Gadella, "Detection of damage in mammalian sperm cells.," *Theriogenology* **65**(5), 958–978 (2006).

20. F. J. Peña et al., "Detection of early changes in sperm membrane integrity pre-freezing can estimate post-thaw quality of boar spermatozoa.," *Anim. Reprod. Sci.* **97**(1-2), 74–83 (2007).

21. H. Rodrı, "Laboratory Semen Assessment and Prediction of Fertility : still Utopia ?* Outcomes from Routine Laboratory Sperm," 312–318 (2003).

22. D. R. Franken and S. Oehninger, "Semen analysis and sperm function testing.," *Asian J. Androl.* **14**(1), 6–13 (2012).

23. T. D. Magrini et al., "Low-level laser therapy on MCF-7 cells: a micro-Fourier transform infrared spectroscopy study.," *J. Biomed. Opt.* **17**(10), 101516 (2012).

24. D. I. Ellis and R. Goodacre, "Metabolic fingerprinting in disease diagnosis: biomedical applications of infrared and Raman spectroscopy.," *Analyst* **131**(8), 875–885 (2006).

25. M. Wojdyr, "Fityk : a general-purpose peak fitting program," *J. Appl. Crystallogr.* **43**(5), 1126–1128, International Union of Crystallography (2010).

26. M. Nichi et al., "Seasonal variation in semen quality in Bos indicus and Bos taurus bulls raised under tropical conditions.," *Theriogenology* **66**(4), 822–828 (2006).

27. M. J. Allen, E. M. Bradbury, and R. Balhorn, "The natural subcellular surface structure of the bovine sperm cell.," *J. Struct. Biol.* **114**(3), 197–208 (1995).

28. Z. Movasaghi, S. Rehman, and D. I. ur Rehman, "Fourier Transform Infrared (FTIR) Spectroscopy of Biological Tissues," *Appl. Spectrosc. Rev.* **43**(2), 134–179 (2008).

29. M. E. Potok et al., "Reprogramming the maternal zebrafish genome after fertilization to match the paternal methylation pattern.," *Cell* **153**(4), 759–772, Elsevier Inc. (2013).

30. M. Benchaib et al., "Influence of global sperm DNA methylation on IVF results.," *Hum. Reprod.* **20**(3), 768–773 (2005).

31. S. Kumar et al., "Atomic force microscopy: a powerful tool for high-resolution imaging of spermatozoa.," *J. Nanobiotechnology* **3**(1), 9 (2005) .

32. K. SAEKI et al., "Fine Surface Structure of Bovine Acrosome-Intact and Reacted Spermatozoa Observed by Atomic Force Microscopy," *J. Reprod. Dev.* **51**(2), 293–298 (2005).

33. P. Fénichel and M. Durand-Clément, "Role of integrins during fertilization in mammals.," *Hum. Reprod.* **13 Suppl 4**, 31–46 (1998).

34. K. L. Asquith et al., "Tyrosine phosphorylation activates surface chaperones facilitating sperm-zona recognition.," *J. Cell Sci.* **117**(Pt 16), 3645–3657 (2004).

35. H. M. Florman and T. Ducibella, *Knobil and Neill's Physiology of Reproduction*, 3rd ed., in *Knobil Neill's Physiol. Reprod.* **1**, 3rd ed., J. D. Neill, Ed., pp. 55–112, Elsevier (2006).

36. R. A. P. Harrison and B. M. Gadella, "Bicarbonate-induced membrane processing in sperm capacitation.," *Theriogenology* **63**(2), 342–351 (2005).